# Metal-Insulator Transition and Orbital Order in PbRuO$_3$

Simon A.J. Kimber[1,2], Jennifer A. Rodgers[2], Hua Wu,[3] Claire A. Murray[2], Dimitri N. Argyriou[1], Andrew N. Fitch[4], Daniel. I. Khomskii[3] and J. Paul Attfield[2 *]

[1]*Hahn-Meitner Institute, 100 Glienicker Straße, 14109 Berlin, Germany*

[2]*School of Chemistry and Centre for Science at Extreme Conditions, University of Edinburgh, King's Buildings, Mayfield Road, Edinburgh EH9 3JZ, UK*

[3]*II. Physikalisches Institut, Universität zu Köln, Zülpicher Straße 77, D-50937 Köln, Germany*

[4]*European Synchrotron Radiation Facility, Boîte Postale 220, 38043 Grenoble Cedex, France*

Anomalous low temperature electronic and structural behaviour has been discovered in PbRuO$_3$. The structure (space group Pnma, a = 5.56314(1), b = 7.86468(1), c = 5.61430(1) Å) and metallic conductivity at 290 K are similar to those of SrRuO$_3$ and other ruthenate perovskites, but a sharp metal-insulator transition at which the resistivity increases by four orders of magnitude is discovered at 90 K. This is accompanied by a first order structural transition to an Imma phase (a = 5.56962(1), b = 7.74550(1), c = 5.66208(1) Å at 25 K) that shows a coupling of Ru$^{4+}$ 4d orbital order to distortions from Pb$^{2+}$ 6s6p orbital hybridization. The Pnma to Imma transition is an unconventional reversal of the group-subgroup symmetry relationship. No long range magnetic order is evident down to 1.5 K. Electronic structure calculations show that hybridization of Pb 6s6p and Ru 4d orbitals and strong spin-orbit coupling stabilise this previously hidden ground state for ruthenate perovskites.

Transition metal oxide perovskites display a remarkable range of electronic phenomena such as superconductivity, colossal magnetoresistances, and coupled charge, orbital and spin orderings [1]. Perovskite-related ruthenates based on low spin $t_{2g}^4$ $Ru^{4+}$ have proved particularly interesting as broad Ru:4d bands lead to metallicity without chemical doping, so very clean itinerant correlated electron physics may be observed in single crystals. The layered perovskite $Sr_2RuO_4$ is a p-wave superconductor [2] while bilayered $Sr_3Ru_2O_7$ shows metamagnetism and non-Fermi liquid behaviour associated with a quantum critical end point at 8 T [3]. All of the cubic-type $ARuO_3$ perovskites (A = Ca, Sr, Ba) are metallic to lowest measured temperatures. $SrRuO_3$ and $BaRuO_3$ are Stoner ferromagnets with Curie temperatures of $T_C$ = 160 and 60 K respectively [4,5,6,7], but $CaRuO_3$ shows non-Fermi liquid behaviour without a magnetic transition [8]. Another ruthenate perovskite, $PbRuO_3$, was synthesised at high pressures in 1970 [9], but the electronic properties of this material have not been reported. $Pb^{2+}$ (ionic radius = 1.49 Å) is intermediate in size between $Sr^{2+}$ (1.44 Å) and $Ba^{2+}$ (1.61 Å) so $PbRuO_3$ is expected to be a ferromagnetic metal with $T_C \approx$ 130 K based on size-controlled bandwidth arguments. However, in this Letter we report that $PbRuO_3$ behaves very differently to the other $ARuO_3$ perovskites, and instead shows a sharp metal-insulator transition at 90 K but without apparent long range spin order. Electron localisation is strongly coupled to the lattice through Ru orbital ordering, resulting in an anomalous structural change to higher lattice symmetry at low temperatures. We propose that strong hybridisation of Ru 4d with Pb 6s and 6p states and spin-orbit coupling are responsible for inducing this alternative ground state for ruthenate perovskites.

Small (ca. 10 mg) polycrystalline pellets of $PbRuO_3$ were synthesised by heating the oxygen deficient pyrochlore $Pb_2Ru_2O_{6.5}$ at 11 GPa and 1100 °C using a Walker type multianvil press. Synchrotron X-ray diffraction profiles in the temperature range 10 < T < 300 K were collected from instrument ID31 at the ESRF, France with wavelength λ =

0.45621 Å, and time-of-flight neutron powder data were recorded using the GEM spectrometer at ISIS, UK over the range 1.5 < T < 300 K. Rietveld fits to diffraction data were performed using the GSAS package [10]. Magnetic susceptibility and electronic resistivity measurements were made using Quantum Design MPMS and PPMS systems.

Good fits to room temperature X-ray ($\chi^2$ = 3.13, $R_{wp}$ = 0.103) and neutron ($\chi^2$ = 1.27, $R_{wp}$ = 0.026) powder diffraction profiles of $PbRuO_3$ were obtained using an orthorhombic Pnma space group model as found for $SrRuO_3$ [11]. We also attempted to fit the data using non-centrosymmetric variants of this structure, in case of a steric ('lone pair') effect from the Pb $6s^2$ state, but no improvements were obtained and the fits diverged. All of the site occupancies refined to within error (ca. 1%) of full occupancy, showing that the sample is stoichiometric [12]. The Pnma superstructure is often observed in oxide perovskites and results from a generic, oxygen-centred, tilting instability rather than specific, transition metal-centred electronic instabilities.

To compare the contributions of $Pb^{2+}$ and $Sr^{2+}$ to the ruthenate band structures, we performed electronic structure calculations for room temperature Pnma type $PbRuO_3$ and $SrRuO_3$ [13] in the local density approximation (LDA) using the full-potential augmented plane wave plus local orbital method [14]. The results (Fig. 1) for $SrRuO_3$ are very similar to those previously reported [5]. The $Sr^{2+}$ states lie far from the Ru 4d - O 2p bands near the Fermi level [6] and even the closest Sr 4d state has negligible hybridization with the Ru 4d states. A large exchange splitting is found and the itinerant ferromagnetic state is stabilised by 25 meV/f.u., in keeping with the 160 K Curie transition. By contrast, the Ru 4d and O 2p states in $PbRuO_3$ lie just between the occupied Pb 6s state and the unoccupied Pb 6p state. The Ru $t_{2g}$ - Pb 6s and Ru $e_g$ - Pb 6p hybridizations, both aided by the O 2p state, and $t_{2g}$-$e_g$ mixing due to the lattice distortion, significantly suppress the exchange splitting of the Ru $t_{2g}$ - O 2p conduction bands, reducing the magnetic stabilization energy to near zero (<4 meV/f.u.).

To confirm that the slight lattice differences between the two materials are unimportant, we also calculated the electronic structure of SrRuO$_3$ using the PbRuO$_3$ parameters and found the electronic and magnetic properties were virtually identical.

An unexpected difference between PbRuO$_3$ and the other perovskite ruthenates was discovered by resistivity measurements (Fig. 2). At ambient temperatures, the PbRuO$_3$ has a resistivity of ~$10^{-1}$ Ωcm with little temperature dependence, characteristic of metallic conduction with a resistive grain boundary contribution. However, on cooling, the resistivity increases sharply by four orders of magnitude (and was immeasurably large below 60 K), signifying a metal-insulator transition at $T_{MI}$ = 90 K. This transition is also evident in magnetic susceptibility data. The high temperature susceptibility $\chi$ is fitted by a combination of Curie-Weiss and Pauli terms ($\chi = C / (T - \theta) + \chi_p$) with $\chi_p$ = 2.09(1) x $10^{-3}$ emu/mol, $C$ = 0.195 emu.K/mol (corresponding to a paramagnetic moment of 1.25 $\mu_B$) and $\theta$ = -54 K. Similar large temperature independent contributions have been reported for non-perovskite ruthenates such as the pyrochlore Tl$_2$Ru$_2$O$_7$ ($\chi_p \approx$ 2 x $10^{-3}$ emu/mol) [15] close to metal-insulator instabilities. A small dip is observed in the magnetic susceptibility on cooling through $T_{MI}$ showing that antiferromagnetic correlations are present in the insulating state. However, no long range magnetic transition is evident down to 4 K although a broad hump with divergence between field and zero-field cooled susceptibilities is observed below 50 K.

A strong coupling of structure to the metal-insulator transition is observed in both X-ray and neutron powder diffraction measurements (Fig. 3). PbRuO$_3$ remains orthorhombic down to 1.5 K, however, the lattice parameters and volume change discontinuously on cooling through the transition (Fig. 4) and surprisingly, the Pnma superstructure reflections with odd (h + k + l) values disappear (see Fig. 3 inset). No new superstructure reflections, peak broadenings or splittings were observed in the low temperature diffraction patterns of PbRuO$_3$, which are indexed by the body-centred space group Imma. This describes another common tilting superstructure of perovskites and a refined Imma model gives excellent fits to both X-ray and neutron data ($\chi^2$= 2.31 and $R_{wp}$ = 0.140; $\chi^2$= 1.77 and $R_{wp}$ = 0.022, respectively for 25 K data sets) [12]. Fits of possible lower-symmetry, acentric body-centred structures were unsuccessful. No magnetic diffraction peaks were observed down to 1.5 K in the GEM time-of-flight neutron diffraction data or in subsequent constant wavelength profiles collected from instrument E6 at the HMI reactor. We estimate the upper limit for any ordered moment to be ~0.5 $\mu_B$.

The (high temperature) Pnma to (low temperature) Imma transition in $PbRuO_3$ is remarkable as Pnma is a subgroup of Imma so a continuous group-subgroup transition from Imma to Pnma is allowed in Landau theory, and is observed in many simple perovskites such as $SrSnO_3$ [16]. The Pnma-Imma 'subgroup-group' transition in $PbRuO_3$ is clearly first order, with a small volume anomaly typical of metal-insulator transitions, and a substantial hysteresis of 20 K in the cell parameters between warming and cooling experiments (Fig. 4) [17]. The 'subgroup-group' structural contribution to the transition entropy is negative, but this is evidently outweighed by the large positive electronic contribution from the delocalisation of Ru 4d electrons.

The evolution of the Ru-O bond distances (Fig. 4) reveals an important aspect of the metal-insulator transition. At room temperature, the $RuO_6$ octahedra are almost regular with Ru-O bond lengths in the range 2.00-2.01 Å, but below $T_{MI}$ a Jahn-Teller distortion is apparent in the Imma structure, with two short Ru-O1 bonds (1.97 Å) aligned approximately along z and four long Ru-O2 bonds (2.02 Å) in the xy plane. To a first approximation, this corresponds to a $d_{xy}^2 d_{xz}^1 d_{yz}^1$ orbital ordering of the $Ru^{4+}$ $t_{2g}^4$ configuration in the insulating Imma phase, creating planes of minority-spin-occupied $d_{xy}$ orbitals, as shown in Fig. 4. $Pb^{2+}$ shows an unusual A-site distortion, having a near-regular square pyramidal coordination with five short Pb-O bonds (Pb-O1, 2.51 Å x 1; Pb-O2, 2.50 Å x 4), while other Pb-O distances are

> 2.82 Å. The O1-Ru-O2 angle of 122.5° shows that this is not a lone pair effect, for which an angle of <90° is expected. The Pb and Ru distortions are cooperative as O1 forms short bonds to Ru and only one short bond to Pb whereas O2 has long bonds to Ru and four short bonds to Pb.

To clarify the orbitally ordered state, we have carried out LSDA+U calculations for the Imma phase with an effective Hubbard U = 3.5 eV. Spin-orbit coupling (SOC) was also included since this ~160 meV interaction is large relative to the calculated crystal field splitting between the $d_{xy}$ and $d_{xz}/d_{yz}$ levels of ~50 meV. Our LSDA+U+SOC calculations gave an insulating ground state with a small gap of ~0.1 eV, verifying that $PbRuO_3$ is in the vicinity of a metal-insulator transition. The minority spin electron has a $0.46(1 + i)d_{xy} - 0.38(1 - i)(d_{xz} + d_{yz})$ orbital state which consists of 42% $d_{xy}$, 29% $d_{xz}$ and 29% $d_{yz}$, with $d_{xy}$ being dominant as expected from the above structural results.

The insulating, orbitally-ordered ground state of $PbRuO_3$ is anomalous in comparison to the other $ARuO_3$ perovskites (A = Ca, Sr, Ba) which remain metallic to lowest temperature. This is not due to a size effect as $CaRuO_3$ ($Ca^{2+}$ radius = 1.34 Å) has the most tilted Pnma superstructure but is stable to orbital order. (The combination of small $Ca^{2+}$ and an imposed tetragonal symmetry in layered $Ca_2RuO_4$ is sufficient to induce a weak orbital order [18], but this phase is a Mott insulator.) The electronic nature of Pb is a key factor and Ru 4d - Pb 6s6p - O 2p hybridizations are evident in the above band structure calculations. This $Pb^{2+}$ covalency is sometimes manifest as a lone pair distortion resulting in ferroelectricity, e.g. in $PbTiO_3$, but lone pair distortions are not observed in either the Pnma or the Imma phases of $PbRuO_3$. Another consequence of covalency is the stabilisation of lower $Pb^{2+}$ coordination numbers than expected from cation size arguments [19] as shown by the change from three short (2.47 Å) Pb-O bonds in the Pnma structure of $PbRuO_3$ to five, described above, in the Imma phase. Hence, the metal-insulator transition in $PbRuO_3$ is driven by electronic instabilities of both cations as the orbital order of $t_{2g}^4$ $Ru^{4+}$ is coupled to an order of $s^2p^0$ $Pb^{2+}$ hybrid states. By contrast, Ru-orbital order is suppressed and the metallic state remains stable in the other $ARuO_3$ perovskites that lack A-cation instabilities.

Orbital order lowers magnetic dimensionality relative to the structural dimensionality and this can open a spin gap in some non-perovskite ruthenates, e.g. a singlet dimerized phase in $La_4Ru_2O_{10}$ [20,21] and possible Haldane chains in $Tl_2Ru_2O_7$ [15]. This seems not to be the case here, however, the observation of a broad susceptibility maximum at 25 K and the lack of a long range magnetic transition down to 1.5 K, which in conjunction with a Weiss temperature of -54 K corresponds to a frustration factor $|\theta/T_c| > 36$, shows that $PbRuO_3$ does not have a conventional ordered magnetic ground state. The divergence of field and zero field cooled susceptibilities evidences some glassy character to the ground state, but there is no obvious source for structural disorder. One possibility is that the combination of orbital order and octahedral tilting (which gives a Ru-O-Ru angle of 159.8°) weakens nearest neighbor antiferromagnetic superexchange interactions in the xy plane so that they become comparable to the next nearest neighbor couplings. This frustrates spin order in the xy plane, leading to one-dimensional (z-direction) magnetic behaviour in the three-dimensional perovskite lattice imposed by orbital order.

In summary, the low temperature properties of $PbRuO_3$ show that normally-hidden orbitally ordered states such as that of degenerate $t_{2g}^4$ $Ru^{4+}$ ions in ruthenate perovskites may be stabilised by coupling to electronic instabilities of other cations. This may provide a strategy for accessing orbitally ordered states of other 4d and 5d transition metal oxide networks. The combined order of Pb s and p hybridized orbitals, Ru d-orbitals and O-centred octahedral tilting instabilities results in an anomalously high symmetry ground state structure that inverts the usual group-subgroup symmetry descent. These distortions also suppress long range spin order in PbRuO$_3$, and further experiments and theoretical work will be needed to elucidate the magnetic ground state.

We acknowledge EPSRC for support and the provision of ESRF and ISIS beamtime and the Leverhulme trust for additional support. HW and DIK are supported by DFG through SFB 608. We thank J.W.G. Bos (Edinburgh); P.G. Radaelli (ISIS); N. Stüer, A. Buchsteiner and D. A. Tennant (HMI) for assistance with diffraction measurements and useful discussions.

* Corresponding author: j.p.attfield@ed.ac.uk

FIG. 1: (color online) Density of states (DOS) of (a) PbRuO$_3$ and (b) SrRuO$_3$ in the Pnma structure calculated by LDA in the ferromagnetic state. Up (down) arrows refers to spin up (down). The Ru t$_{2g}$ DOS is scaled by 1/3 and the Sr DOS is scaled by 1/2. The exchange splitting of the Ru t$_{2g}$ - O 2p bands is significantly suppressed in PbRuO$_3$.

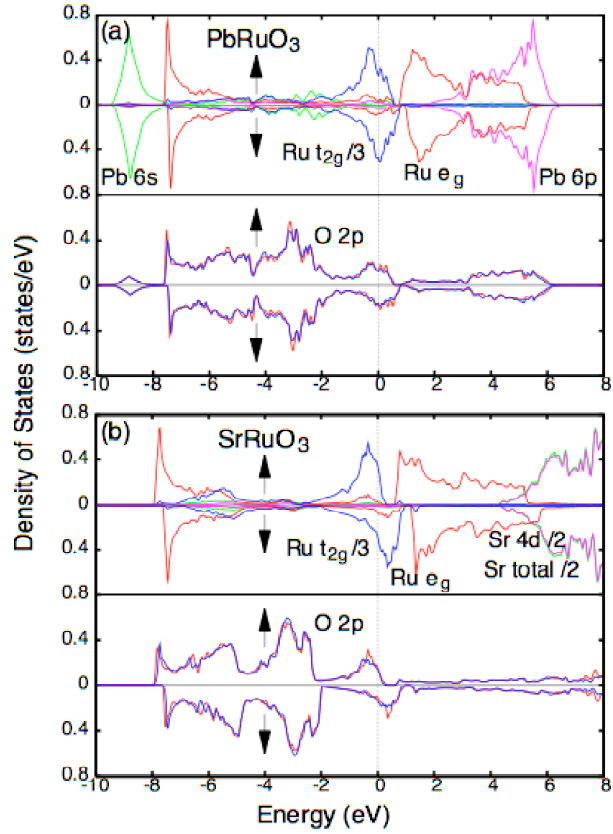

FIG. 2: (color online) Resistivity variation for PbRuO$_3$, inset shows magnetic susceptibility measured in a 1 T field with a modified Curie-Weiss fit as described in the text.

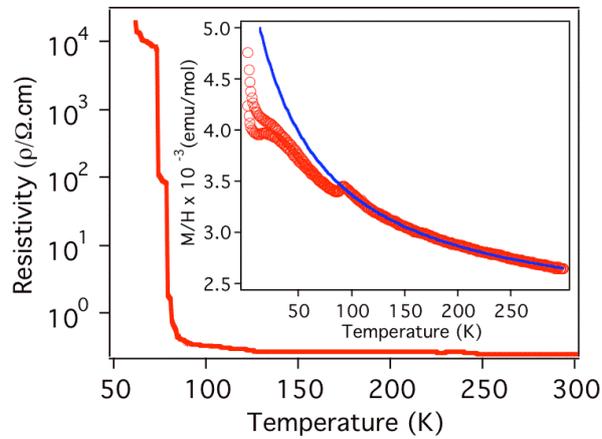

FIG. 3: (color online) Fit of the Imma model to the time-of-flight neutron diffraction profile of PbRuO$_3$ at 1.5 K. Inset shows the disappearance of Pnma X-ray superstructure reflections as the Imma phase is formed on cooling.

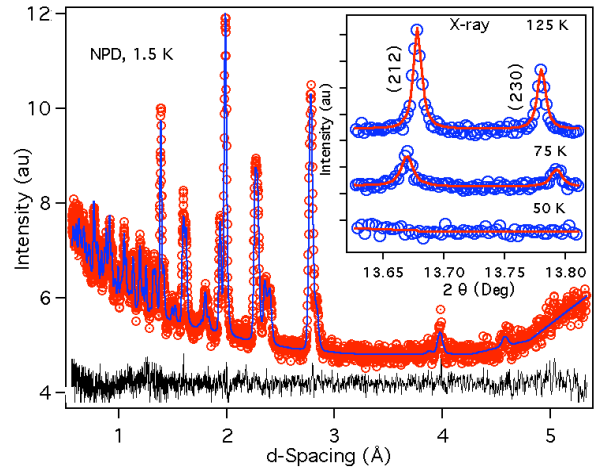

FIG. 4: (color online) Cell volume (top) and Ru-O distance (bottom) variations for PbRuO$_3$ from fits to neutron diffraction data. Upper inset shows hysteresis in the lattice parameters, normalised to the cubic perovskite cell, from X-ray measurements while warming or cooling at 10 K/min. Idealised $d_{xy}$ orbital order in the Imma phase is shown on the lower panel.

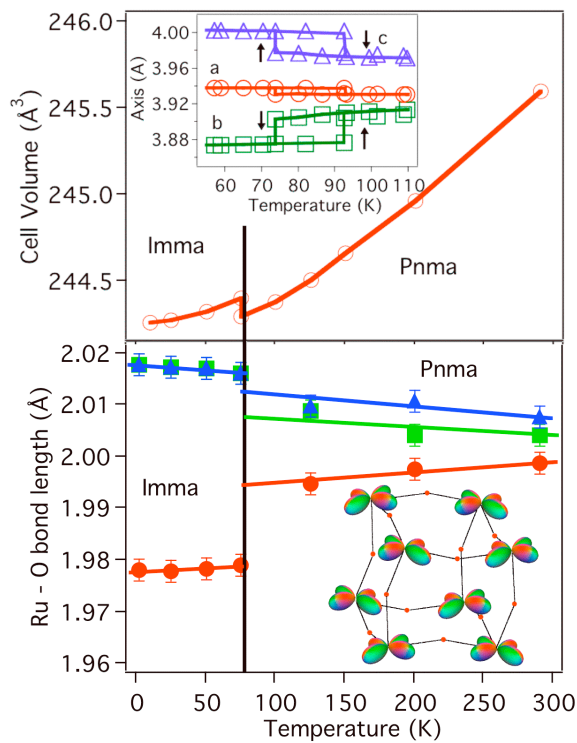